# Model for triggering of non-volcanic tremor by earthquakes

Naum I. Gershenzon[1,2], Gust Bambakidis[1]

[1]Physics Department, Wright State University, 3640 Colonel Glenn Highway Dayton, OH 45435
[2]Department of Earth and Environmental Sciences, Wright State University, 3640 Colonel Glenn Highway  Dayton, OH 45435

[1] There is evidence of tremor triggering by seismic waves emanating from distant large earthquakes. The frequency content of both triggered and ambient tremor are largely identical, suggesting that this property does not depend directly on the nature of the source. We show here that the model of plate dynamics developed earlier by us is an appropriate tool for describing tremor triggering. In the framework of this model, tremor is an internal response of a fault to a failure triggered by external disturbances. The model predicts generation of radiation in a frequency range defined by the fault parameters. Thus, although the amplitude and duration of a tremor burst may reflect the "personality" of the source, the frequency content does not. The model also explains why a tremor has no clear impulsive phase, in contrast to earthquakes. The relationship between tremor and low frequency earthquakes is discussed.

**Introduction**
[2] Deep non-volcanic tremor arises inside of, or in close proximity to, well-developed subduction and transform faults at a certain depth ranges. The spatio-temporal distribution of tremor reflects faults geodynamics and could be used for monitoring the latter. It has been observed that (1) bursts of tremor accompany slip pulses in so-called Episodic Tremor and Slip (ETS) phenomena [e.g. *Rogers, & Dragert*, 2003; *Obara*, 2009];  (2) seismic waves from either the local medium or from distant large earthquakes can trigger tremor [*Obara*, 2003; *Rubinstein J. et al* 2007, 2009; *Peng et al*, 2009; *Miyazawa & Mori*, 2006; *Miyazawa & Brodsky*, 2008; *Fry et al*, 2011]; (3) the intensity of tremor varies with tidal stress [*Rubinstein J. et al*, 2008; *Nakata et al*, 2008; *Thomas et al*, 2009; *Lambert et al*, 2009]. While the duration and amplitude of a tremor burst varies depending on the source, the spectral composition remains essentially the same. The question arises as to how various external stress disturbances, spanning a wide ranges of amplitudes and frequencies, can all trigger tremor in the 2 to 30 Hz range in the fault area.
[3] It has been shown that tremor is triggered and modulated by Rayleigh wave [*Miyazawa & Mori*, 2006; *Miyazawa & Brodsky*, 2008; *Fry et al*, 2011] as well as by Love wave [*Rubinstein et al*, 2007; 2009; *Peng et al*, 2009] from distant large earthquakes. While large amplitude and proper direction of the wave are necessary conditions for tremor triggering, these are not the only conditions required. Triggered tremor appears to be adjacent to an area of ongoing SSE [*Fry et al*, 2011] or (in the case of short SSE) coincides with the location of the SSE source area [e.g., *Hirose and Obara*, 2006; *Gomberg et al*, 2010], suggesting that the condition that the fault be close to failure is also necessary. Triggered tremor usually appears in the same areas as ambient

tremor, with the same frequencies and polarizations [*Rubinstein et al*, 2009; *Peng et al*, 2009]. Overall comparison of different characteristics of ambient and triggered tremor suggests that they are generated by the same physical process [e.g. *Rubinstein et al*, 2010].

[4] Recently we developed a Frenkel-Kontorova (FK)-type model, which describes quantitatively the dynamic frictional process between two surfaces [*Gershenzon et al*, 2009; *Gershenzon et al*, 2011; *Gershenzon & Bambakidis,*2011]. Predictions of the model are in agreement with laboratory frictional experiments [*Rubinstein S. et al,* 2004; *Ben-David et al*, 2010]. This model has also been applied to describe tremor migration patterns in ETS phenomena as well as the scaling law of slow slip events [*Gershenzon et al*, 2011]. In the continuum limit, the FK model is described by the nonlinear sine-Gordon (SG) equation. The basic solutions of the latter are kinks and phonons [e.g. *McLauglin & Scott*, 1978] which, in our context, may be interpreted as slip pulses and radiation respectively. In the framework of the model, radiation may arise due to a variety of mechanisms such as acceleration/deceleration of a slip pulse, interaction of a slip pulse with large asperities, and the action of an external stress disturbance on the frictional interface. The first two mechanisms may be used to describe generation of tremor during ETS events and will be considered in detail in a future publication. In this Letter we will focus on the latter mechanism.

[5] Here is our suggested scenario. The low frequency Rayleigh and/or Love wave generated by a distant earthquake increases the tangential stress and/or decreases the effective normal stress in the vicinity of a fault, so the Coulomb stress temporarily increases, hence decreasing static friction. There are always spots within a fault with residual tangential stress. Such spots may remain, for example, after a slip pulse passes the region. Thus, a seismic wave with sufficiently large amplitude and proper direction may trigger local failure (slip), exciting a radiation mode inside the fault. Then the radiation (as a small-amplitude, localized relative motion of plate surfaces with zero net slip) propagates along the fault attenuated due to friction and geometrical spreading. Since the fault is immersed in a 3D solid body, the radiation inside the fault will generate *S* waves (tremor) propagating as far as the Earth's surface. It is important to note that the frequency of these waves is defined by the radiation frequency, hence by the fault parameters, and does not depend on the frequency of the external source.

**Model**

[6] It has been shown that the dynamics of a frictional surface may be described by the SG equation (see [*Gershenzon et al*, 2009; *Gershenzon & Bambakidis,* 2011] for justification):

$$\frac{\partial^2 u}{\partial t^2} - \frac{\partial^2 u}{\partial x^2} + \sin(u) = \Sigma_S^0 - f, \qquad (1)$$

where $u(x,t)$ is the relative shift of the frictional surfaces at time $t$ and distance $x$ in the slip direction and $\Sigma_S^0(x,t)$ and $f(x,t)$ are the external shear stress and frictional force per unit area; $u$, $x$ and $t$ are in units of $b/(2\pi)$, $b/A$ and $b/(cA)$, respectively, where $b$ is a typical distance between asperities, $c^2 = c_l^2(1-2\nu)/(1-\nu)^2$, $c_l$ is the longitudinal acoustic velocity (or *P* wave velocity), and ν is Poisson's ratio; $\Sigma_S^0$ and $f$ are both in units of $\mu A/(2\pi)$, where $\mu$ is the shear modulus. The parameter $A \approx \Sigma_N/\sigma_p$, where $\Sigma_N$ is the effective normal stress (normal stress minus fluid pressure) and $\sigma_p$ is the penetration hardness. The variables $\varepsilon = \sigma_s = \partial u/\partial x$, and $w = \partial u/\partial t$ are interpreted as the dimensionless strain, stress and slip velocity in units of $A/(2\pi)$, $\mu A/\pi$ and $cA/(2\pi)$, respectively.

[7] The basic solutions of equation (1) are kinks (solitons), breathers and phonon radiation. The nonlinear dispersion equation for the radiation is [*McLauglin & Scott*, 1978]

$$\omega^2 - k^2 = \frac{\pi^2}{4K[(1-\cos a)^{1/2}/2]^2}$$

where ω is the angular frequency, $k$ is the wave number, $a$ is the wave amplitude and $K$ is the complete elliptic integral of the first kind. Here we will consider the case of small amplitude ($a<<\pi$), i.e. a wave with amplitude much smaller than $b$. In this case the dispersion relation is simplified:

$$\omega^2 - k^2 = 1 \tag{2}$$

and the group velocity of the wave packet is

$$V = d\omega/dk = k/(1+k^2)^{1/2} \tag{3}$$

Two important conclusions follow from equations (2) and (3) for values of $k<<1$: (1) the angular frequency is almost constant and approximately equal to unity (ω≈1), which means that the frequency (in dimensional units) $f = (cA/b)\cdot\omega/(2\pi) \approx cA/(2\pi b)$ is defined by the fault parameters and effective normal stress only and does not depend on the parameters of a particular source and (2) the group velocity $V \approx k <<1$, thus the wave packet can propagate along a fault with velocity much smaller than seismic velocities.

[8] Let's assume that there is an area on the fault close to failure, i.e. an area with residual localized shear stress almost equal to the value of the static frictional force per unit area. This island of residual stress may have remained after passage of a slip pulse due to the presence of an asperity with size larger than the typical size of surrounding asperities. Suppose that a seismic wave of large amplitude arrives at this spot increasing the tangential stress and/or decreasing the normal stress in such a way that locally and temporally the tangential stress exceeds static friction. The ensuing failure (slip) overcomes or destroys this large asperity, producing a regular or slow earthquake or simply aseismic slip. As we will see shortly this impulse may also produce radiation inside the fault. Let's suppose, for simplicity, that passage of the impulse past this spot reduces the right hand side of equation (1) to $\Sigma_S^0 - f = \sigma_0 \cdot \delta(x) \cdot \delta(t)$, where δ is the Dirac delta-function and $\sigma_0$ is the strength of the disturbance. Using the results of a perturbation analysis of SG equation [*McLauglin & Scott*, 1978] we find that the radiation field produced by the perturbation is

$$u(x,t) = \frac{a_w}{2\pi}\int_{-\infty}^{\infty}\frac{1}{\sqrt{1+k^2}}\sin(t\cdot\sqrt{1+k^2})\exp(-ikx)dk \tag{4}$$

Figure 1 show the results of a numerical integration of equation (4) for $\sigma_0 = 1$. One can see that the disturbance (as a small relative shift of plates) originating at point x=0 at time t=0 propagates along a fault in both directions with unit velocity (velocity *c* in dimensional units) (Figure 1 a-d). The wave number *k* ranges from large values close to the wave fronts to small values close to the center. The value of *k* in the center decreases in time and becomes much less than unity when t>>2π. In this case ω≈1 (see equation (2) and Figure 1e), so after a short time the frequency of the radiation in close proximity to the center reaches the value f=ω/(2π)≈1/(2π) and does not change much thereafter. The oscillation frequency at points close to the fronts is higher (see Figure 1f-h). Therefore the angular frequency ranges from low (ω=1) in the center to higher frequencies $\omega = (1 + k_{init}^2)^{1/2}$ at the periphery of the disturbance, where $k_{init}$ is the characteristic wave number of the initial disturbance (see Figures 1e-h). The periodic localized oscillations of the plate surface generate *S*-type seismic waves with the same mix of frequencies; the latter may propagate trough the crust to the Earth's surface. Thus the model predicts generation of tremor in the ω range from 1 to $\omega = (1 + k_{init}^2)^{1/2}$. Oscillations close to the center contribute to the lower portion of the tremor frequency range and, respectively, oscillations at the periphery of the disturbance to the higher portion.

**Results**
[9] Our model is described by two adjustable parameters: the typical distance *b* between asperities and the dimensionless parameter *A*. In a previous article by Gershenzon et al [2011] we showed that a slip pulse in an ETS event could be represented as a solitonic solution of equation (1). Then the parameter *b* should be equal to the typical slip produced by one ETS event, i.e. b≈30 mm. The parameter *A* is the ratio between the effective normal stress and the penetration hardness [*Gershenzon & Bambakidis,* 2011], both of which are unknown in our case. We can estimate the value of *A* using the central frequency of tremor. Supposing that this frequency corresponds to the dimensionless frequency f=1/(2π) of our model we can find *A* from the relation $f = cA/(2\pi b)$ yielding $A \approx 1.5 \cdot 10^{-4}$ if f=4 Hz and c = 5 km/s. Having the values of these two adjustable parameters as determined from experimentally well-defined quantities, we now use our model to calculate the values of other parameters which are difficult to measure.
[10] Supposing that σ$_p$≈0.018μ(1+ν) [*Rabinowicz,* 1965] and taking μ=30 GPa we estimate the value of the effective normal stress Σ$_N$≈ *A*·σ$_p$ to be 0.1 MPa. Although this estimate is not particularly accurate, due to uncertainty of the σ$_p$ value, it nevertheless allows us to conclude that the effective normal stress is quite low, i.e. a high fluid pressure is required for the model to work. Using equation (4) and supposing that tremor is triggered by an local failure of amplitude 10 mm we may calculate the radiation parameters, i.e. the maximal amplitudes of the slip $a_{rad}$ (not the initial failure slip, but a slip produced by radiation), slip velocity $a_v$ and shear stress $a_\sigma$; these values are $a_{rad}$ =1 mm, $a_v$ =2.4 mm/s and $a_\sigma$=0.3 MPa, respectively. It is also useful to estimate the minimal size ($L_{min}$) of the frictional area which can emit tremor. Since *k* should

satisfy the condition $k^2 \ll 1$, the size $L$ of the disturbance should be larger than $L_{min}=2\pi$ or, in dimensional units, $L \gg L_{min} = 2\pi b/A \approx 1.2$ km.

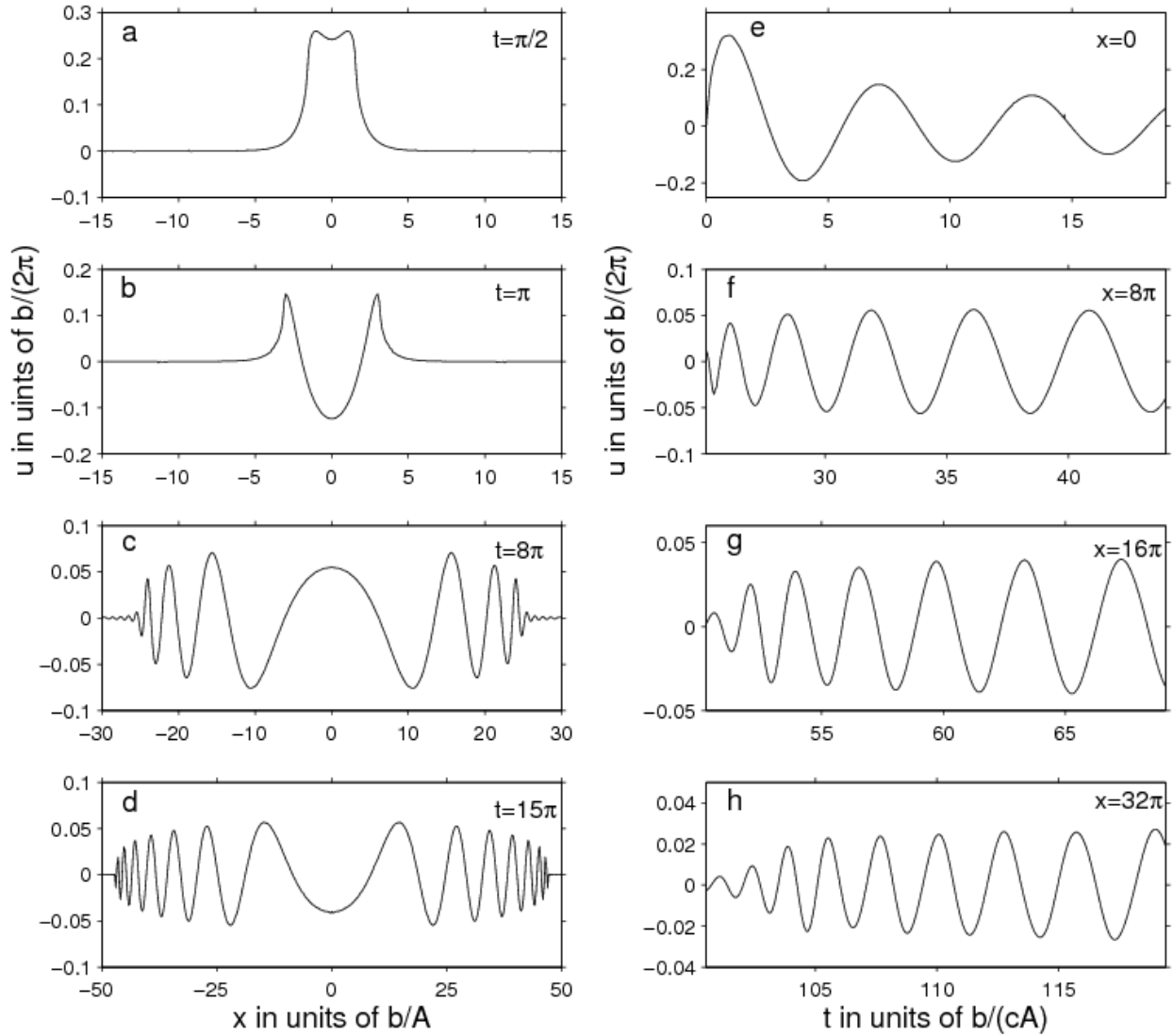

Figure 1. The evolution of the shift disturbance in space (along a fault) for various moments of time (panels on the left) and in time for various distances from the center (panels on the right). Disturbance originates at point x=0 and time t=0 by the external source $\delta(x)\cdot\delta(t)$ and propagates in both directions. The local wave number at any particular time decreases going from the center to the wave fronts (Figure 1c and d). The period of oscillation at point $x=0$ approaches $2\pi$ ($f \approx 1/(2\pi)$) after a short time ($t>2\pi$) from the beginning (see Figure 1e). The oscillation period at points $x=8\pi, 16\pi, 32\pi$ progressively decreases compared with the value at point $x=0$ (see Figure 1f-h). The disturbance at the center plays the dominant role in the frequency distribution of the emitted tremor; the peripheral disturbance contribute to its high frequency range.

[11] In the scenario proposed the radiation gradually increases in time. The typical time of tremor development is $2\pi$ (dimensionless units). For times $t>2\pi$ the wave number around the point $x = 0$ gradually decreases, thus the typical size of the disturbance in the center (the first oscillation about the center) gradually increases (see Figure 1 b-d). For lager times ($t >> 2\pi$) the size of the "central disturbance" increases very slowly since the group velocity $V$ of the wave packet becomes smaller and smaller as $k$ decreases in time ($V \approx k$). This may explain a major difference between tremor and earthquakes. The latter have a clear impulsive phase, reflecting the propagation of rupture with velocity approaching seismic velocities. In the case of tremor generated by the radiation mode of a frictional surface, the initial phase should gradually increase in time, and this is actually observed. The typical time for tremor to develop can be estimated from the relation $T = L_{min}/V \approx 2\pi/k^2$ (using formula (3) with $k^2<<1$) or, in dimensional units, $T \approx (2\pi/k^2) \cdot (b/(cA)) \approx 2.5\,\text{s}$ if $k^2=0.1$. So tremor can develop over a time span of up to a few seconds.

[12] Based on analysis of tremor duration-amplitude scaling, Watanabe et al [2007] concluded that the size of the source of tremor should be "scale-bound rather than scale-invariant". This result is consistent with the prediction of our model. Indeed, whatever the source exciting a local shear failure, the latter disturbance will quickly grow in size, but the "central disturbance" will grow only up to a size of $2L_{min}$, after which the expansion rate will drop considerably in time (see formula (3)). Since tremor is supposedly generated within and in close proximity to this central disturbance, this implies that for given values of the parameters $b$ and $A$ the size of the source should be scale-bound.

[13] It has been observed that (1) a considerable portion of tremor signals include superposed LFE waveforms [*Shelly et al*, 2007(a); *Brown et al*, 2009] and (2) signals from VLF earthquakes are usually buried in tremor signals [*Ito et al*, 2007; 2009]. What is the relationship between tremor and LFEs and between tremor and VLFs? Our model implies that any failure inside a fault excites a radiation mode. Under appropriate conditions this radiation may propagate along the fault and produce tremor by mechanisms described above. In this respect all types of earthquakes, including LFE and VLF, can be sources of tremor. However, if the size of the earthquake source equals the size of the failure area, the size of the tremor source is much larger even though it originates in the same place. This may be another reason why source tremor is hard to determine.

[15] There have been a few attempts to evaluate the effective seismic moment of tremor [*Kao et al*, 2010; *Fletcher & McGarr*, 2011]. As already pointed out, tremor does not represent a rupture itself, as in an earthquake, but rather reflects an oscillatory disturbance on the frictional surface. Such oscillatory disturbances may exist under specific conditions intrinsic to the depth and fluid content of the fault, and may be excited by a rupture observed, for example, as a LFE. In this context the seismic moment, which depends on slip and slip area, may not much reflect the physics of the process.

[16] Two mechanisms have been proposed to explain the appearance of tremor [e.g. *Rubinstein et al*, 2010]: (1) tremor is generated by fluid-flow processes; (2) tremor reflects a frictional process (rupture) on a fault with low rupture speeds. Both approaches require the presence of fluid and high fluid pressure. Our model does not directly require the presence of fluid in the source area, but to make it consistent with observed tremor parameters, we require a comparatively small effective normal stress, which may not be expected at such depths without the presence of high pressure fluid. The specific features predicted by our model are (1) the size of the emitted area is on the order of a few km; (2) tremor accompanies aseismic slip and earthquakes including LFE and VLF; (3) there is no particular dependence of tremor characteristics on the frequency of the triggering seismic wave.